\documentclass[preprint,aps,showpacs]{revtex4}

\usepackage{graphicx}
\usepackage{dcolumn}
\usepackage{bm}
\newcommand{\be}{\begin{equation}}
\newcommand{\ee}{\end{equation}}

\newcommand{\fig}[1]{Fig. (\ref{#1})} 
\newcommand{\eqa}{\begin{eqnarray}}
\newcommand{\eeq}{\end{eqnarray}}  
\begin{document}
\preprint{Preprint for New J. Phys. 2008}

\title{3D Electron Fluid Turbulence at Nanoscales in Dense Plasmas}
\author{Dastgeer Shaikh} 
\email{das0007@uah.edu}
\affiliation{
Center for Space Plasma and Aeronomic Research\\
The University of Alabama in Huntsville\\
Huntsville. Alabama, USA 35899}
\author{P. K. Shukla}
\email{ps@tp4.rub.de}
\affiliation{Institut f\"ur Theoretische Physik IV, Ruhr-Universit\"at Bochum, 
D-44780 Bochum, Germany\\
Department of Physics, Ume\aa~ University, SE-90187 Ume\aa,~ Sweden\\
SUPA Department of Physics, University of Strathclyde, Glasgow G4 ONG, Scotland\\
Instituto de Plasmas e Fus\~ao Nuclear, Instituto Superior T\'ecnico, Universidade
T\'ecnica de Lisboa, 1049-001 Lisboa, Portugal}
\received{31 March 2008}
\revised{8 July 2008}

\begin{abstract}
We have performed three dimensional nonlinear fluid simulations of
electron fluid turbulence at nanoscales in an unmagnetized warm dense
plasma in which mode coupling between wave function and electrostatic
potential associated with underlying electron plasma oscillations
(EPOs) lead to nonlinear cascades in inertial range. While the wave
function cascades towards smaller length scales, electrostatic
potential follows an inverse cascade.  We find from our simulations
that quantum diffraction effect associated with a Bohm potential plays
a critical role in determining the inertial range turbulent spectrum
and the subsequent transport level exhibited by the 3D EPOs.

\end{abstract}
\pacs{52.27.Gr,52.35.Ra,71.10Ca}                             
\maketitle


\section{introduction}
Studies of collective phenomena at nanoscales in dense matters are of
great importance in diverse areas of physics, including the fields of
plasmonics \cite{r1,r1a,r2,r3,r3a}, semiconductors \cite{r4},
nano-electromechanical systems \cite{r5}, quantum-diodes \cite{r6},
nanotubes and nanowires \cite{r7}, quantum free electron lasers
\cite{r7a}, as well as astrophysical bodies \cite{r7b,r7c} and intense
laser-solid density plasma interaction experiments \cite{r8} for x-ray
and $\gamma$-ray sources.  In dense plasmas, the electrons are highly
degenerate and quantum mechanical effects (e.g. electron tunneling
arising from the finite width of the electron wave function) play an
essential role at nanoscales.  Since degenerate electrons follow the
Fermi-Dirac statistics, there appear new electron equation of state
and new forces involving the quantum Bohm potential \cite{r9} and
electron-$1/2$ spin effects \cite{r10} in dense quantum plasmas. It
then turns out that due to the intrinsic nonlinearities associated
with the Fermi pressure law and quantum forces, there exists
possibility of localizing electrostatic \cite{r11,r12} and
electromagnetic \cite{r13} wave energies at nanoscales in dense
quantum plasmas. Here we report simulation studies of three-dimensional (3D)
electron fluid turbulence at nanoscales in an unmagnetized warm dense
plasma.  It is found that 3D nonlinearly interacting electron plasma
oscillations (EPOs) \cite{r14} in a dense quantum plasma exhibit
nano-structures and associated energy spectra that are markedly
different from those reported earlier for the 2D case \cite{r12}.
Furthermore, we stress that the present 3D turbulence properties
of electron plasma oscillations in our dense quantum plasma are
significantly different from those in a classical plasma 
\cite{zakharov,goldman,sulem}. In the latter, strong electron plasma wave 
turbulence has been studied by invoking parametric interactions \cite{murtaza} 
and by using either multi-dimensional cubic nonlinear Schr\"odinger equation \cite{goldman,sulem} 
or Zakharov equations \cite{zakharov}, which are different from our 3D coupled 
nonlinear Schr\"odinger-Poisson equations in dense quantum plasmas.

The surge for studying numerous nonlinear processes in dense quantum
plasmas lies in a hope to transfer information through localized
nano-structures one is able to create and sustain in plasmas. The
present work dealing with 3D electron fluid turbulence shares a great
deal of knowledge with classical fluid turbulence \cite{r15,r16},
plasma turbulence \cite{r17,r18}, and superfluid turbulence involving
the Bose-Einstein condensates (BECs) \cite{r19,r20} in ultracold
gases. Both in fluids and plasmas as well as in BECs, one encounters
the phenomena of inverse energy cascades in which energy transfer from
small scales sustains large scale circulations/structures in the flow,
and results in a steady-state inertial range with power-law scaling,
as was originally predicted by Kolmogorov, Kraichnan and Iroshnik
\cite{r15,r16}.  while the dynamical equations depicting inverse
cascades in fluids and plasmas are the Navier-Stokes and
Charney-Hasegawa-Mima equations \cite{r17,r18}, the energy cascade
scenario in BECS is described by the Gross-Pitaevskii equation
\cite{r19,r20}. In Sec II, we describe model equation governing the
dynamical evolution of 3D dense fermi quantum plasma. We also present
conservation laws admitted by the set of 3D equations. In Sec III,
nonlinear 3D simulation results describing turbulence in such system
are described. Mode structures and corresponding Kolmogorov-like
spectra are also discussed. Turbulence transport is described in Sec
IV, and finally the conclusions are contained in Sec V.

\section{Model equations}
In dense quantum plasmas, the Wigner-Poisson (WP) model has been used
to derive a set of quantum hydrodynamic (QHD) equations \cite{r21} in
the mean field approximation.  The QHD equations include the continuity, 
momentum and Poisson equations. The quantum nature appears in the electron 
momentum equation through the pressure term, which requires knowledge of 
the Wigner distribution for a quantum mixture of electron wave functions, 
each characterized by an occupation probability satisfying the Pauli exclusion 
principle. The quantum part of the electron pressure is represented as a quantum
force $-\nabla \phi_B$, where the Bohm potential is \cite{r9} $\phi_B
= - (\hbar^2/2m_e \sqrt{n_e})\nabla^2 \sqrt{n_e}$. Here $\hbar$ is the
Planck constant divided by $2\pi$, $m_e$ is the electron mass, and
$n_e$ is the electron number density. Defining the effective wave
function $\psi =\sqrt{n_e({\bf r},t)} \exp[iS({\bf r}, t)/\hbar]$,
where $\nabla S({\bf r}, t) = m_e {\bf u}_e({\bf r},t)$ and ${\bf u}_e
({\bf r}, t)$ is the electron velocity, the electron momentum equation
can be represented as an effective nonlinear Schr\"odinger (NLS)
equation \cite{r23}, in which there appears a coupling between the
wave function and the electrostatic potential associated with the
EPOs. The electrostatic potential is determined from the Poisson
equation.  We thus have the coupled NLS and Poisson equations, which
govern the dynamics of nonlinearly interacting EPOs is a warm dense 
quantum plasma.

In this paper, we carry out simulations of 3D NLS and Poisson equations 
in order to understand the properties of 3D electron fluid turbulence
(involving nano-structures and associated electron transport) in a warm 
dense plasma. We find that nonlinear couplings between different scales
EPOs are responsible for creating small-scale electron density clumps, 
while the electrostatic potential assumes large-scale structures. The total 
energy associated with our 3D electron fluid turbulence at nanoscales 
processes a characteristic spectrum which is a {\it non-} Kolmogorov-like. 

For our 3D electron fluid turbulence studies, we use the NLS--Poisson 
equations \cite{r11,r21}
\begin{equation} 
i \sqrt{2H} \frac{\partial \Psi}{\partial t}+ H \nabla^2\Psi 
+ \varphi \Psi - |\Psi|^{4/3}\Psi = 0,
\end{equation}
and 
\begin{equation}
\nabla^2\varphi = |\Psi|^2-1, 
\end{equation}
which are valid at zero electron temperature for the Fermi-Dirac
equilibrium distribution. In Eqs. (1) and (2) the wave function 
$\Psi$ is normalized by $\sqrt{n_0}$, the electrostatic
potential $\varphi$ by $k_B T_F/e$, the time $t$ by the electron
plasma period $\Omega_{pe}^{-1}$, and the space ${\bf r}$ by the Fermi
Debye radius $\lambda_D$. We have introduced the notations $\lambda_D=
(\epsilon_0 k_B T_F/n_0 e^2)^{1/2} \equiv V_F/\Omega_{pe}$ and $\sqrt{H} =
\hbar \omega_{pe}/\sqrt{2} k_B T_F$, where $\epsilon_0$ is the  
electric permittivity, $k_B$ is the Boltzmann constant and the Fermi electron temperature 
$k_B T_F = (\hbar^2/2m_e)(3\pi^2)^{1/3} n_0^{2/3}$, $e$ is magnitude of the
electron charge, and $\Omega_{pe} =(n_0e^2/\epsilon_0 m_e)^{1/2}$ is the
unperturbed electron plasma frequency. The origin of the various terms 
in Eq.(1) is obvious. The first term is due to the electron inertia, 
the $H$-term is associated from the quantum diffraction effect involving 
the Bohm potential, $\varphi \Psi$ comes from the nonlinear coupling
between the scalar potential (associated with the space charge electric 
field resulting form oscillations of the electrons around immobile ions)
and the electron wave function, and the fourth  term in the left-hand side 
of (1) is the contribution of the 3D electron pressure ($p_e 
=m_eV_{F}^2 n_e^{5/3}/5n_0^{2/3}$) for the Fermi plasma
with a quantum statistical equation of state.

Equations (1) and (2) admit a set of conserved quantities \cite{r8}:
the number of electrons $N= \int \Psi^2 dV$, the electron
momentum ${\bf P} =- i \int \Psi^\ast \nabla \Psi dV$, the electron
angular momentum ${\bf L}=- i \int \Psi^\ast {\bf r} \times \nabla
\Psi dV$, and the total energy ${\cal E}= \int [-\Psi^\ast H
\nabla^2 \Psi + |\nabla\varphi|^2/2 +(3/5)|\Psi|^{10/3}] dV$,
where $dV =dx dy dz$.  In obtaining the total energy ${\cal E}$, we 
used the relation $\partial {\bf E}/\partial t= i H(\Psi \nabla \Psi^\ast-\Psi^\ast
\nabla \Psi)$, where the electric field ${\bf E}=- \nabla\varphi$. The
conserved quantities are used to maintain the accuracy of the numerical
integration of Eqs. (1) and (2). We note that linearizing the latter 
one obtains the EPO frequency $\omega = \left(\Omega_{pe}^2 + k^2 V_F^2 
+ \hbar^2 k^4/4 m_e^2\right)^{1/2}$, which exhibits the dispersive behavior
of the EPOs. In the short wavelength regime characterized by 
$k^2 \gg 4 m_e^2 V_F^2/\hbar^2$, one notices that the dispersion associated with
electron tunneling effect dominates over that involving the quantum statistical 
electron pressure.  

\section{nonlinear 3D  simulations of quantum plasmas}

We have developed 3D fluid code to investigate nonlinear interactions
between multi-scales EPOs described by (1) and (2). Our 3D fluid code
is based on Fourier expansion of the bases using a fully de-aliased
pseudospectral numerical scheme \cite{r22}. The nonlinear
de-convolution of Fourier modes is performed by computing the
nonlinear triad interactions $\tilde{f}({\bf x},t) \tilde{g}({\bf
  x},t) = \sum_{{\bf k} ={\bf k}'-{\bf k}''} f({\bf k}',t)g({\bf
  k}'',t) \delta({\bf k}'-{\bf k}'')$ at each time, which survive for
only those coupled modes which satisfy the Fourier triad constraint
${\bf k}={\bf k}'-{\bf k}''$.  These nonlinear interactions in Fourier
space follow from a Kolmogorov phenomenology of spectral energy
transfer, and are mediated predominantly by the neighboring
modes. Furthermore, such nonlinear interactions in the local spectral
space conserve constants of motion of Eqs. (1) and (2), as presented
above.  The temporal integration is performed by 4th order Runge Kutta
method.  The spectral distribution for turbulent fluctuations is
initialized isotropically (no mean fields are assumed) with random
phases and amplitudes in Fourier space.  The evolution variables use
periodic boundary conditions. The initial isotropic turbulent spectrum
was chosen close to $k^{-2}$, with random phases in all three
directions.  The choice of such (or even a flatter than $-2$) spectrum
treats the turbulent fluctuations on an equal footing and avoids any
influence on the dynamical evolution that may be due to the initial
spectral non-symmetry.  Note, however, that the local as well as
global mean flows may subsequently be generated by self-consistently
excited nonlinear instabilities. Finally, the algorithm employed in
our 3D fluid code ensures conservation of total energy and mean fluid
density per unit time in the absence of charge exchange and external
random forcing, and it is massively parallelized using Message Passing
Interface (MPI) libraries to facilitate higher resolution. The
conserved quantities, described in section II, are used to monitor the
numerical accuracy of our 3D quantum fluid plasma code. Time step
during the numerical integration varies between $10^{-3}$ and
$10^{-4}$. The code uses usual 2/3 de-aliasing of the Fourier modes,
such that the largest $k$ is determined typically by
$k_{max}=(2/3)(1/2)N$, where $N$ is the total number of modes in each
direction.  The time step used in our 3D simulations not only
preserves the conserved quantities, but it also minimizes the aliasing
errors.

\begin{figure}[ht]
\centering
\includegraphics[width=12.cm]{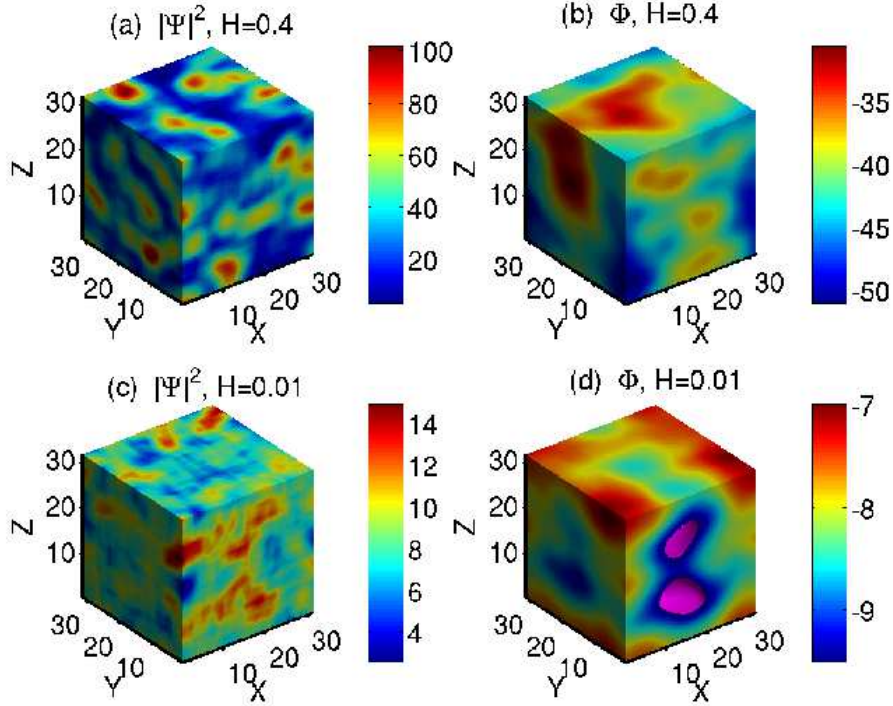}
\caption{\label{fig1} Fluctuations in the electron number density resulting
  from a steady state turbulence simulations of our 3D electron
  quantum plasma. Forward cascades are responsible for the generation
  of relatively small-scale fluctuations in a decaying 3D electron
  quantum plasma as shown in (a). Large scale electron flow structures are
  present in the electrostatic potential, essentially resulting from a
  merging of smaller scale fluctuations as shown in (b). The latter is
  known as the inverse cascade process. $H=0.4$ is used in (a) and
  (b). Figures (c) and (d) depict the electron number density and the electrostatic
  potential for $H=0.01$.}
\end{figure}

The localized initial turbulent spectral distribution, concentrated 
at the lower wavenumbers, evolves in time following 3D nonlinear
electron plasma wave interactions. Since the initial energy is
localized in the large scale fluctuations, the latter drive turbulent
processes through migration of energy towards relatively small scales
Consequently, larger eddies transfer their energy to smaller ones
through a forward cascade.  During the forward cascade process, each
Fourier mode in the inertial range spectrum obeys the vector triad
constraints \cite{murtaza} imposed by the vector relation ${\bf k} + {\bf p} = 
{\bf q}$. These nonlinear interactions involve the neighboring Fourier 
components (${\bf k},{\bf p},{\bf q}$) that are excited in the local inertial
range turbulence. We have performed a number of simulations to verify
the consistency of our results in a strong turbulence regime.  In
our 3D simulations, we have explored two dense plasma systems that are
characterized by different physical parameters, {\it viz.}  dense
warm plasmas in the next generation laser-based plasma compression (LBPC)
schemes \cite{r23}, and the superdense astrophysical bodies
\cite{r24}(e.g. interior of white dwarf stars). It is expected that in 
LBPC schemes, the electron number density may reach $10^{33}$ m$^{-3}$ 
and beyond. Hence, we have $\Omega_{pe} =1.76 \times 10^{18}$ s$^{-1}$,
$k_B T_F = 1.7 \times 10^{-16}$ J, $\hbar \Omega_{pe} = 1.7 \times
10^{-16}$ J, and $H = 1$.  The Fermi Debye radius $\lambda_D = 0.1$
$\AA$.  On the other hand, in the interior of white dwarf stars, we
typically have $n_0 \sim 10^{36}$ m$^{-3}$ (such values are also
common in dense neutron stars and supernovae), yielding $\Omega_{pe}
=5.64 \times 10^{19}$ s$^{-1}$, $k_B T_F = 1.7 \times 10^{-14}$ J,
$\hbar \Omega_{pe} = 5.64 \times 10^{-15}$ J, $H \approx 0.3$, and
$\lambda_D = 0.025$ $\AA$.  The numerical solutions of Eqs. (1) and
(2) for $H=0.4$ and $H=0.025$ (corresponding to $n_0 =10^{33}$
m$^{-3}$ and $n_0 =10^{36}$ m$^{-3}$, respectively) are displayed in
Fig. (1), which are the electron number density (the left figure) and
electrostatic (ES) potential distributions (the right figure) in the
$(x,y,z)$-cube.

One of most notable features of our simulations encompassing 3D
electron fluid turbulence is that it exhibits a dual cascade phenomenon, 
as shown in Fig. 1.  The electron density distribution in Fig 1 shows a 
tendency to generate smaller length-scale structures, while the ES potential 
cascades towards larger scales. The co-existence of small and larger scale
structures in turbulence is a ubiquitous feature of various 2D and 3D
turbulence systems. For example, in 2D hydrodynamic turbulence, the
incompressible fluid admits two invariants, namely the energy and the
mean squared vorticity.  The two invariants, under the action of an
external forcing, cascade simultaneously in turbulence, thereby
leading to a dual cascade phenomena. On the other hand, 3D MHD
turbulence exhibits forward cascades of energy and an inverse cascade
of magnetic helicity. In these processes, the energy cascades towards
smaller length-scales, while the magnetic helicity in MHD transfers
spectral power towards larger length-scales. By contrast, the fluid
vorticity in 3D hydrodynamics is prohibited from an inverse cascade.
The randomly excited 3D Fourier modes nonetheless transfer the
spectral energy by conserving the constants of motion in $k$-space. 
In freely decaying quantum electron fluid turbulence reported here, 
the energy contained in the large-scale eddies is transferred to the 
smaller scales, leading to a statistically stationary inertial regime 
associated with the forward cascades of one of the invariants. 
Decaying turbulence often leads to the formation of coherent structures 
as turbulence relaxes, thus making nonlinear interactions rather inefficient 
when they are saturated.  It is to be noted further that the long scale flow
generation in our 3D simulations is observed to be directly
proportional to the parameter $H$.  Intermittent flows are thus
generated for a small value of $H$, while strong and large scale
flows in the ES potential are formed when the magnitude of $H$ is large
(see, e.g. \fig{fig1}). The physical basis of such observation can be
elucidated from the following arguments. The parameter $H$, which is
the ratio between the energy density of the EPOs and the electron
kinetic energy density of a warm dense quantum plasma, is associated 
with a diffraction-like term in Eq. (1) i.e. $H\nabla^2\Psi$. In this term, 
the negative imaginary part of the complex evolutionary variable $\Psi$
essentially determines the rate of dissipation corresponding to the
smaller scales. The smaller is $H$, more the dissipation is
concentrated at the smaller scales and vice versa. For a moderately
higher magnitude of the $H$ parameter, there exists a strong tendency 
in EPO's to dissipate the smaller and intermittent turbulent eddies.  
It is therefore this $H$ parameter which essentially characterizes 
electron flows at nanoscales in our 3D simulations.

\begin{figure}
\centering
\includegraphics[width=12.cm]{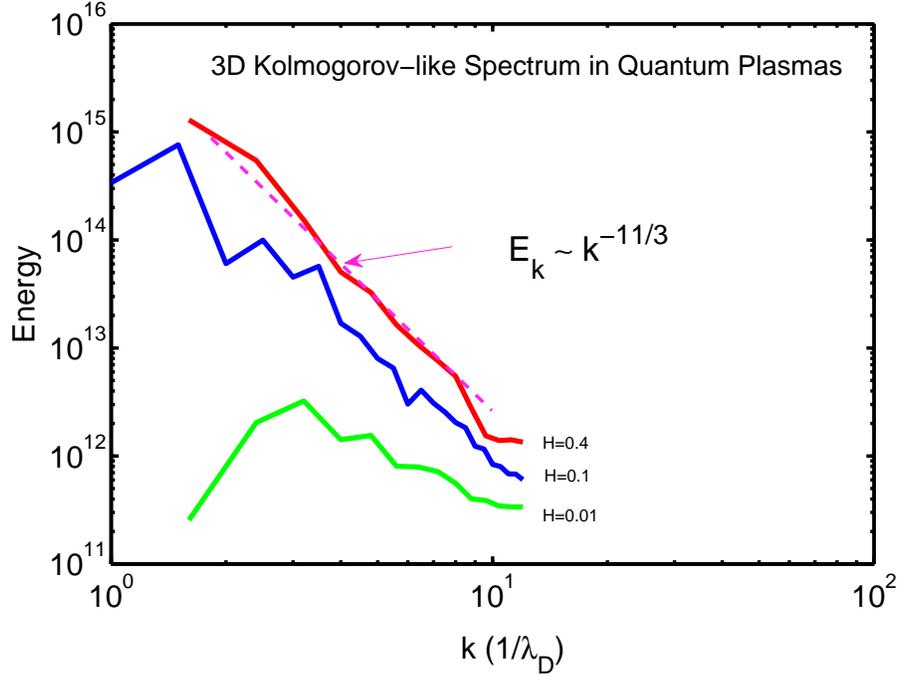}
\caption{\label{fig2} Energy spectrum of 3D EPOs in a
  forward cascade regime. A Kolmogorov-like $k^{-11/3}$ has been
  observed for $H=0.4$. The spectral index changes with respect to the
  parameter $H$. Numerical resolution is $128^3$.}
\end{figure}

While the power spectrum for nonlinear EPOs exhibits an interesting
feature in our 3D simulations, its scaling is not universal and is
determined critically by the parameter $H$. For instance, we find a 3D
Kolmogorov-like power spectrum $k^{-11/3}$ in some range of $H$ values
as shown in \fig{fig2}.  The corresponding omnidirectional spectrum
thus exhibits a $k^{-5/3}$ scaling.  Spectral index nevertheless
changes with $H$, as noted also in the study of 2D fluid turbulence
\cite{r12}. However, the spectral slope in the latter was found to be
close to the Iroshnikov-Kraichnan (IK) power law \cite{r25,r26}
$k^{-3/2}$, rather than the usual Kolomogrov scaling \cite{r27}
$k^{-5/3}$. The origin of the differences in the observed spectral
indices resides with the nonlinear character of the underlying warm
dense plasmas, as nonlinear interactions in 2D and 3D systems are
governed typically by different nonlinear forces. The latter modify
the spectral evolution of turbulent cascades to a significant degree.

\begin{figure}
\centering
\includegraphics[width=12.cm]{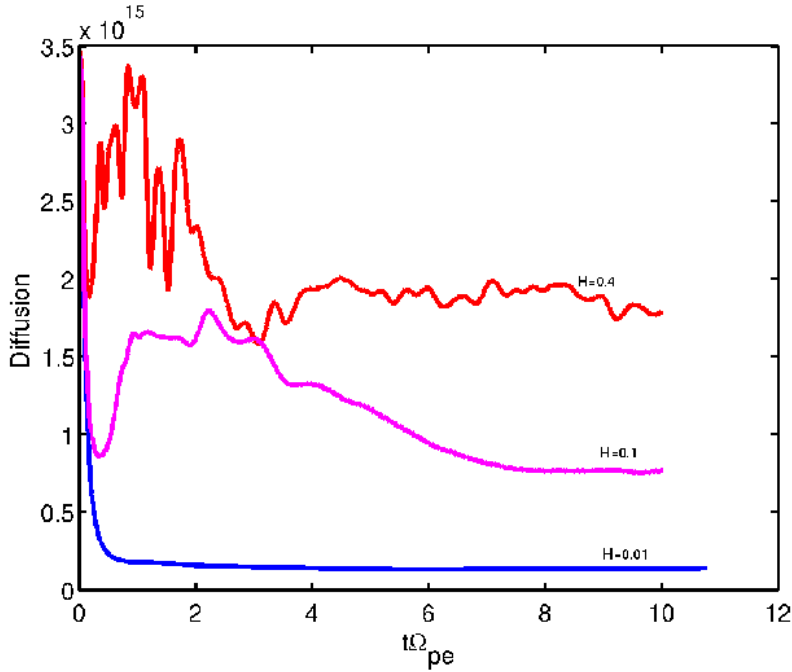}
\caption{\label{fig3} Time evolution of effective electron
  diffusion coefficients measured for different values of
  $H$. Interestingly, smaller values of $H$ corresponds to an
  effective low diffusion coefficient. The latter characterizes
 the  presence of small scale turbulent eddies which results in
  suppression of the electron transport.}
\end{figure}

\section{Electron transport caused by turbulent fields}

We finally study the electron diffusion coefficient in the presence of
small and large scale turbulent EPOs. The effective electron diffusion
coefficient produced by the momentum transfer can be calculated from
$D_{eff} = \int_0^\infty \langle {\bf P}({\bf r}, t) \cdot {\bf P}
({\bf r}, t + t^\prime) \rangle dt^\prime$, where ${\bf P}$ is the electron
momentum and the angular bracket denotes spatial averages and the
ensemble averages are normalized to unit mass. The effective diffusion
coefficient $D_{eff}$, resulting from 3D structures, essentially
relates the diffusion processes associated with random translational
motions of electrons in nonlinear fields of localized EPOs.  
It is {\it remarkable} to note that the electron transport can be effectively
suppressed when the magnitude of the parameter $H$ is decreased. This is
shown in \fig{fig3}.  The outcome of \fig{fig3} is contrary to our 2D
results \cite{r12}, where the effective electron diffusion was lower
when the field perturbations were Gaussian and increases later rapidly
with the eventual formation of longer length-scale structures. Unlike
the 2D case, the electron diffusion in 3D case is suppressed eventually 
because of the presence of small scale turbulent fluctuations. However, 
the steady state diffusion coefficient increases with the increase
of $H$. This can be understood as follows. Higher the $H$
valuer, the stronger is the small scale damping. When the small
scale fluctuations are smeared out, the dynamical evolution of 3D
quantum electron plasma wave turbulence is predominantly governed by
the large scale flows, which consequently lead to a higher level of
transport. This finding is further consistent with the fact that the
parameter $H$ dictates the characteristic evolution properties of 
the 3D EPOs, as described above. 

\section{Summary}
In conclusion, we have presented results from 3D
nonlinear fluid simulations of the electron fluid turbulence at nanoscales
in an unmagnetized warm dense plasma. The mode couplings between the
electron wave function and the electrostatic potential associated with
the underlying electron plasma oscillations (EPOs) lead to the onset of
nonlinear interactions and subsequent cascades in the inertial range.
We find from our 3D simulations that the dispersion effect
associated with the quantum Bohm potential plays a critical role in determining
the inertial range turbulent spectrum and the subsequent transport
level exhibited by the 3D EPOs. For instance, a Kolmogorov-like
$k^{-11/3}$ 3D spectrum is observed for $H=0.4$, whereas the spectrum
flattens out for a smaller value of $H$. Correspondingly, the
electron transport is higher for the higher $H$ values. Finally, the wave
function cascades towards smaller length scales, while the
electrostatic potential follows an inverse cascade.  We reiterate that
the present investigation of 3D EPO turbulence is a necessary
prerequisite for understanding the complex dynamical phenomena occurring
at nanoscales in warm dense plasmas, such as those in laboratory and
astrophysical settings.


\begin{thebibliography}{}
\bibitem{r1}
W. L. Barnes, A. Dereux, and T. W. Ebbesen, Surface plasmon subwavelength
optics. {\it Nature} (London) {\bf 424}, 824 (2003).
\bibitem{r1a}
M. Marklund, G. Brodin, L. Stenflo, and C. S. Liu, New quantum limits 
in plasmonic devices. {\it arXiv:0712.3145v1} (2007).
\bibitem{r2}
S. A. Maier, {\sl Plasmonics} (Springer, Berlin, 2007).
\bibitem{r3}
D. E. Chang, A. S. Sorensen, E. A. Demler, and M. D. Lukin, A single-photon 
transistor using nano-scale surface plasmons. {\it Nature Phys.} {\bf 3}, 807 (2007).
\bibitem{r3a}
J. M. Pitarke, V. M. Silkhon, E. V. Chulikov, and P. M. Echenique, Theory of 
surface plasmons and surface-plasmon polaritons. {\it Rep. Prog. Phys.} {\bf 70},
1 (2007).
\bibitem{r4}
P. A. Markowich, C. A. Ringhofer, and C. Schmeiser, {\sl Semiconductor Equations}
(Springer, Berlin, 1990).
\bibitem{r5}
H. C. Craighead, Nanoelectromechanical systems. {\it Science} {\bf 290}, 1532 (2000).
\bibitem{r6}
L. K. Ang and P. Zhang, Ultrashort-pulse Child-Langmuir law in the 
quantum and relativistic regimes. {\it Phys. Rev. Lett.} {\bf 98}, 164802 (2007).
\bibitem{r7}
M. Lee, J. Im, B. Y. Lee, S. Myung, J. Kang, L. Huang, Y. K. Kwon, and S. Hong,
Linker-free directed assembly of high performance integrated devices
based on nanotubes and nanowires. {\it Nature Nanotech.} (London) {\bf 1}, 66 (2006).
\bibitem{r7a}
A. Serbeto, J. T. Mendon\c{c}a, K. H. Tsui, and R. Bonifacio, Quantum wave kinetics
of high-gain free electron lasers. {\it Phys. Plasmas} {\bf 15}, 013110 (2008). 
\bibitem{r7b}
G. Chabrier, D. Saumon, and A. Y. Potekhin, Dense plasmas in astrophysics: from giant
planets to neutron stars. {\it J. Phys. A: Math. Gen.} {\bf 39}, 4411 (2006).
\bibitem{r7c}
A. K. Harding and D. Lai, Physics of strongly magnetized neutron stars. 
{\it Rep. Prog.  Phys.} {\bf 69}, 2631 (2006).
\bibitem{r8}
S. H. Glenzer, O. L. Landen, P. Neumayer, R. W. Lee, K. Widmann, S. W. Pollaine,
R. J. Wallace, G. Gregori, A. H\"oll, T. Bornath, R. Thiele, V. Schwarz, W.-D. Kraeft,
and R. Redmer, Observation of plasmons in warm dense matter. {\it Phys. Rev. Lett.} {\bf 98}, 
065002 (2007).
\bibitem{r9}
C. L. Gardner and C. Ringhofer, Smooth quantum potential for the hydrodynamic model.
{\it Phys. Rev. E} {\bf 53}, 157 (1996).
\bibitem{r10}
M. Marklund and G. Brodin, dynamics of spin-$1/2$ quantum plasmas. 
{\it Phys. Rev. Lett.} {\bf 98}, 025001 (2007).
\bibitem{r11}
P. K. Shukla and B. Eliasson, Formation and dynamics of dark solitons and 
vortices in quantum electron plasmas. {\it Phys. Rev. Lett.} {\bf 96}, 245001 (2006).
\bibitem{r12}
D. Shaikh and P. K. Shukla, Fluid turbulence in quantum plasmas. {\it Phys. Rev. Lett.} {\bf 99},
125002 (2007).
\bibitem{r13}
P. K. Shukla and B. Eliasson, Nonlinear interactions between electromagnetic waves and electron 
plasma oscillations in quantum plasmas. {\it Phys. Rev. Lett.} {\bf 96}, 096401 (2007).
\bibitem{r14}
D. Pines, Quantum plasma physics. {\it J. Nucl. Energy: Part C: Plasma Phys.} {\bf 2}, 5 (1961).
\bibitem{zakharov}
V. E. Zakharov, Collapse of Langmuir Waves. {\it Sov. Phys. JETP} {\bf 62}, 908 (1972).
\bibitem{goldman}
M. V. Goldman, Strong turbulence of plasma waves. {\it Rev. Mod. Phys.} {\bf 66}, 709 (1984).
\bibitem{sulem}
C. Sulem and P. L. Sulem, {\sl The Nonlinear Schro\"dinger Equations} (Springer, New York, 1999),
Chap. 13.
\bibitem{murtaza}
G. Murtaza and P. K. Shukla, Nonlinear generation of electromagnetic waves in a magnetoplasma.
{\it J. Plasma Phys.} {\bf 31}, 423 (1984).
\bibitem{r15}
M. Lesieur,  {\sl Turbulence in Fluids} (Kluwer, Dordrecht, 1990).
\bibitem{r16}
U. Frisch, {\sl Turbulence} (Cambridge University Press, Cambridge, England, 1995).
\bibitem{r17}
W. Horton and A. Hasegawa, Quasi-two-dimensional dynamics of plasmas and fluid.
{\it Chaos} {\bf 4}, 227 (2004).  
\bibitem{r18}
A. Hasegawa, Self-organization processes in continuous media. {\it Advances in Physics}
{\bf 34}, 1 (1985).
\bibitem{r19}
M. Kobayashi and M. Tsubota, Kolmogorov spectrum of superfluid turbulence: Numerical
analysis of the Gross-Pitaevskii equation with a small dissipation. 
{\it Phys. Rev. Lett.} {\bf 94}, 065302 (2005).
\bibitem{r20}
M. Kobayashi and M. Tsubota, Thermal dissipation in quantum turbulence. 
{\it Phys. Rev. Lett.} {\bf 97}, 145301 (2006).
\bibitem{r21} 
G. Manfredi, How to model quantum plasmas. {\it Fields Inst. Commun.} {\bf 46}, 263 (2005).
\bibitem{r22}
D. Gottlieb and  S. A. Orszag, {\sl Numerical Analysis of Spectral Methods} (SIAM, Philadelphia, 1977).
\bibitem{r23}
V. M. Malkin, N. J. Fisch, and J. S. Wurtle, Compression of powerful x-ray pulses to attosecond 
by stimulated Raman backscattering in plasmas. {\it Phys. Rev. E} {\bf 75}, 026404 (2007).
\bibitem{r24}
A. K. Harding and D. Lai, Physics of strongly magnetized neutron stars. {\it Rep. Prog. Phys.}
{\bf 69}, 2631 (2006).
\bibitem{r25}
A. N. Kolmogorov, The local structure of turbulence in incompressible viscous fluid for 
very large Reynolds' numbers. {\it C. R. Acad. Sci. USSR} {\bf 30}, 301 (1941).
\bibitem{r26}
P. Iroshnikov, Turbulence of a conducting liquid in a strong magnetic field. 
{\it Sov. Astron.} {\bf 7}, 566 (1963).
\bibitem{r27} 
R. H. Kraichnan, Inertial-range spectrum of hydromagnetic turbulence. 
{\it Phys. Fluids} {\bf 8}, 1385 (1965).
\end{thebibliography}
\end{document}